\documentclass[aps,pra,showpacs,superscriptaddress,twocolumn,floatfix,nofootinbib]{revtex4}
\usepackage{amsmath}    
\usepackage{graphicx}   
\usepackage{verbatim}   
\usepackage{color}      
\usepackage{subfigure}  
\usepackage{hyperref}   
\usepackage[bbgreekl]{mathbbol}
\usepackage{amsfonts}
\newcommand{\vct}[1]{{\bf #1}}
\usepackage{subfigure}
\renewcommand\Re{\operatorname{Re}}
\renewcommand\Im{\operatorname{Im}}

\begin{document}
\title{Casimir forces between cylinders at different temperatures}

\author{Vladyslav~A. Golyk}
\affiliation{Massachusetts Institute of Technology, Department of
  Physics, Cambridge, Massachusetts 02139, USA}
\author{Matthias Kr\"uger}
\affiliation{Massachusetts Institute of Technology, Department of
  Physics, Cambridge, Massachusetts 02139, USA}
\author{M. T. Homer Reid}
\affiliation{Massachusetts Institute of Technology, Department of
 Physics, Cambridge, Massachusetts 02139, USA}
\affiliation{Research Laboratory of Electronics, Massachusetts Institute of Technology, Cambridge MA 02139, USA}
\author{Mehran Kardar}
\affiliation{Massachusetts Institute of Technology, Department of
 Physics, Cambridge, Massachusetts 02139, USA}

\begin{abstract}

We study Casimir interactions between cylinders in thermal non-equilibrium,
where the objects as well as the environment are held at different temperatures.
We provide the general formula for the force, in a one reflection approximation,
for cylinders of arbitrary radii and optical properties.
As is the case for equilibrium, we find that the force for optically diluted cylinders
can be obtained by appropriate summation of the corresponding result for spheres.
We find that the non-equilibrium forces are generally larger than their equilibrium counterpart
at separations greater than the  thermal wavelength.
They may also exhibit oscillations as function of separation, leading to stable points of zero net force.
These effects are particularly pronounced for thin conducting cylinders
(e.g. 40nm diameter nano-wires of tungsten) due to their large emissivity.
\end{abstract}


\maketitle
\section{Introduction \& Summary}
There has been considerable recent activity in fabrication of micro and nano electromechanical
devices which actuate forces in response to currents.
At sub-micron scales classical electromechanical forces are supplanted by
forces  due to quantum and thermal fluctuations~\cite{most97}.
In equilibrium these are manifested as van der Waals and Casimir forces~\cite{Parsegian},
while the passage of currents may well lead to temperature imbalances resulting
in thermal radiation and non-equilibrium forces.
Fluctuation-induced forces are typically non additive, and (especially for conductors) with
sensitive dependence on shape and material properties.
Nanowires and nanotubes provide a simple example of an extended shape, with
several possible applications {~\cite{app1,app2,app3,emig11}}.
Moreover, cylindrical geometries are amenable to high precision experiments {~\cite{onofrio1,onofrio2,onofrio3,onofrio4,onofrio5,Decca11}},
providing good contrast to the more widely studied spherical geometry~\cite{Mohideen}.
Cylinders can also be easier fixed positionally and held at different temperatures for possible
experimental study of non-equilibrium effects~\cite{singer}.

In his seminal paper~\cite{Casimir48} Casimir calculated the force between two perfectly conducting parallel
plates, due to the quantum zero point fluctuations of the electromagnetic (EM) field in the intervening vacuum.
Lifshitz~\cite{Lifshitz56} later extended Casimir's computations to the case of real dielectric materials and to
finite temperatures.
A key step in Lifshitz's approach is to include thermal and quantum fluctuations of currents (sources)
in the dielectric, following the formalism of fluctuational  electrodynamics pioneered by Rytov~\cite{Rytov}.
Typically at small separations zero-point fluctuations shape the force, whereas at separations large compared to the thermal wavelength $\lambda_T$, thermal effects dominate~\cite{Lifshitz56,milonni94,mkepl}.
Rytov's formalism is also appropriate to out of equilibrium steady states in which each object is
held at a different temperature. There is extensive literature on the topic of non-equilibrium interactions between two atoms (or molecules)~\cite{Kweon93,Power94,Cohen03,sherk1, Rodriguez10, sherk2, sherk3}. Recently out of equilibrium Casimir forces have been considered in several systems:
parallel plates~\cite{Antezza08,ante11,dean11}, modulated plates~\cite{Bimonte09}, plate and an atom in different setups~\cite{Hankel02,Antezza05,Ellingsen10,ante11,ante12}, two spheres and sphere and a plate~\cite{mkepl}.  {Formalisms for arbitrary objects were presented in Refs.~\cite{mkprl,ante11}.} A common feature for setups involving compact objects is the need to account for the contribution of
the environment to the force, which depends on a possibly different ambient temperature.

In this paper, we consider non-equilibrium forces between two parallel
cylinders characterized by an arbitrary  (including axially anisotropic) dielectric function.
Using the Rytov formalism  general expressions are obtained for forces between wires maintained at
different temperatures from each other and the environment.
These expressions are then analytically and numerically studied in a number of cases.
With the general reader in mind, in the remainder of this section we provide both an outline of the
paper as well as a summary of its main results.
The interested reader can then proceed to the detailed derivations that follow in subsequent sections.

Analytical results are presented in Sec.~\ref{Analytic}, starting with a brief introduction to
the Rytov formalism in \ref{formalism}.
The fluctuating EM field in the space between the cylinders is sourced by fluctuating currents in the
cylinder, with an additional contribution from the surrounding.
The current correlations in each source can be related by a fluctuation dissipation theorem to the
corresponding temperature~\cite{Rytov}; the stress tensor (from which forces are computed)
is then related (via Green's functions) to the source fluctuations.
The force on each cylinder now has two contributions: an interaction force sourced by the other
cylinder, and a self-force due to modification of the EM fields sourced by itself through influences of the other cylinder.
These two contributions are computed respectively in sections \ref{Fint} and \ref{Fself},
within a so-called {\em one reflection approximation} in which multiple
scatterings of the EM field sourced by either cylinder are ignored.
The results are further simplified for thin cylinders, with radii $R_1$ and $R_2$  small compared to thermal wavelength, relevant skin-depth and separation $d$.
In this limit, the non-equilibrium force per unit length is proportional to  $R_1^2 R_2^2$, and decays at large
separation as $d^{-1}$.
It is well known that in the optically dilute limit, with dielectric response $ {\varepsilon}\to 1$, {\em equilibrium} Casimir
forces become pair-wise additive, obtained by summing over contributions of pairs of polarizable volume elements.
In Sec.~\ref{dilute} we show that a corresponding result holds out of equilibrium, namely that the
force on cylinders of sufficiently diluted media can be obtained by pairwise addition of corresponding forces
for chains of small spheres~\cite{mkepl} by treating a thin cylinder as a chain of small spheres.

\begin{figure}\centering
\subfigure[]{
\includegraphics[width=7 cm]{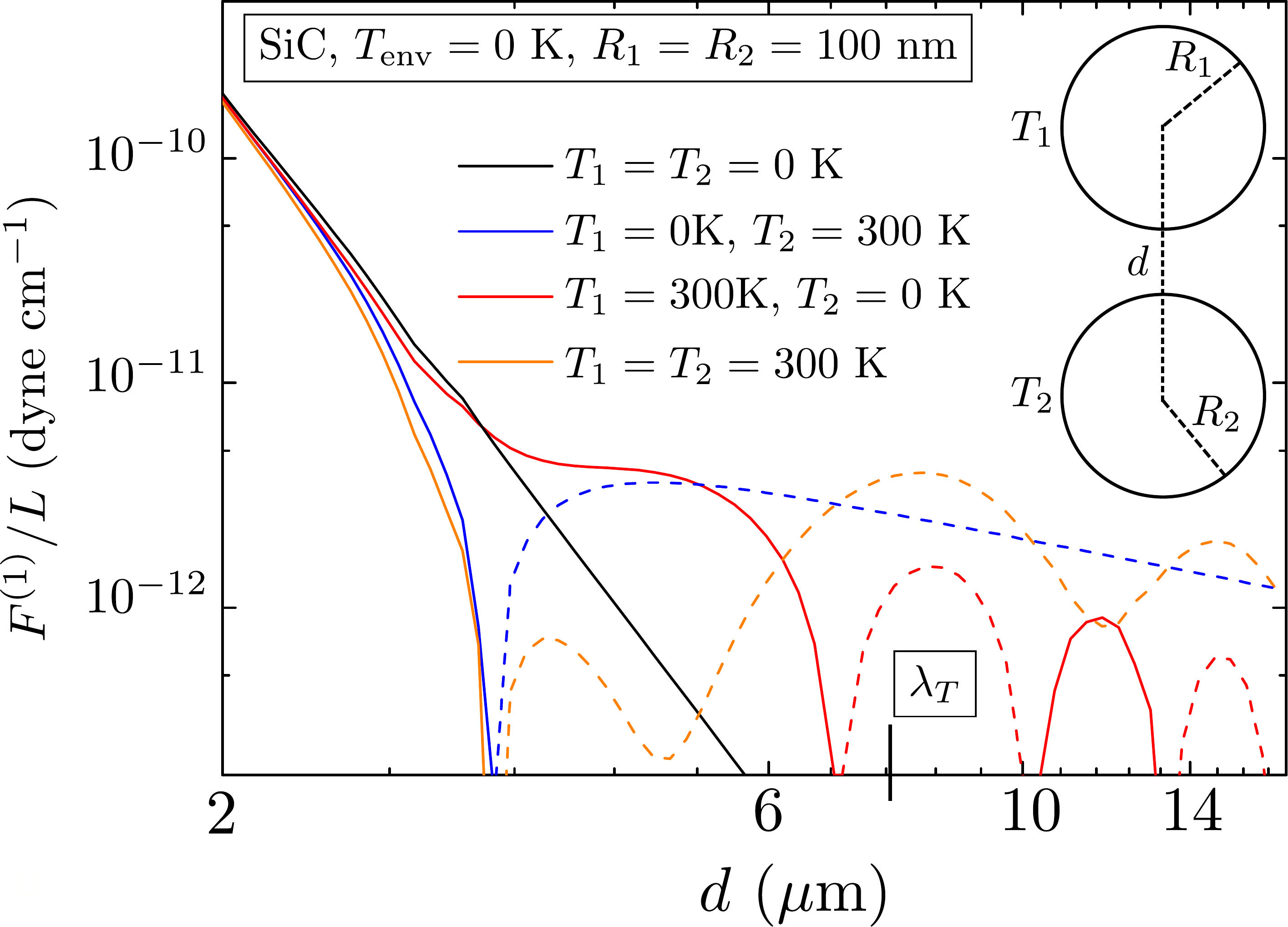}
\label{SiC0}} \subfigure[]{
\includegraphics[width=7 cm]{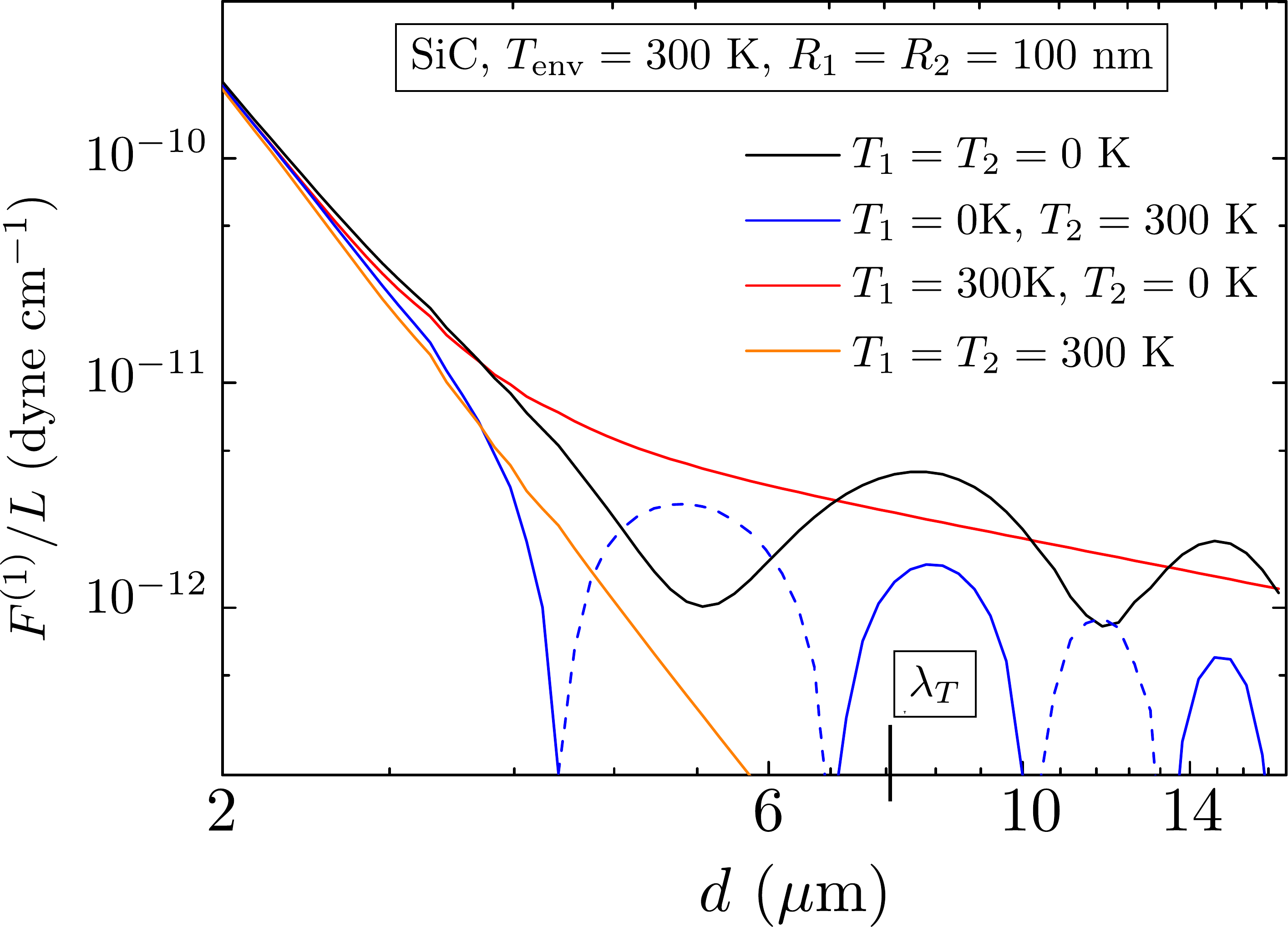}
\label{SiC300}} \caption{(color online). Total force on cylinder 1 per unit length in a system of two SiC cylinders with equal radii $R=0.1\mu$m
at separation $d$ in a a) cold (0 K) b) warm (300K) environment. Dashed lines indicate repulsion. Points of change from repulsive to attractive force with increasing $d$ correspond to stable points of zero force.}\label{SiC}
\end{figure}

As discussed in Sec.~\ref{numerics}, precise calculations of the force can be performed numerically
for different materials, starting from optical data in the form of a (frequency dependent) dielectric
response $ {\varepsilon}(\omega)$.
In particular, we report results for silicon-carbide (insulator) and tungsten (metal):
Figure~\ref{SiC} depicts forces for two SiC nanowires of radii $R_1=R_2=0.1\mu$m; each wire or the environment
is at 0K or 300K, for a total of 8 possible combinations. The top panel correspond to the four cases
where the environment is at 0K, the bottom to where it is at 300K ambient temperature.
To depict results on a logarithmic scale, attractive forces are indicated by solid lines and repulsive forces by
dashed lines.
The equilibrium force (whether at 0K or 300K) is attractive at all separations $d$;
non-equilibrium forces deviate strongly from the equilibrium force at distances larger than the
thermal wavelength $\lambda_T(300K)\approx 7.6\mu$m.
The most significant deviation is for the force on a cold cylinder due to a hot cylinder at $T_{\rm env}=0K$ (top panel)
which becomes repulsive due to radiation pressure, falling off as $1/d$ at large separations.
Note that for the same setup the force experienced by the hot cylinder is not equal and opposite,
instead undergoing  oscillations at a wavelength related to the resonance of SiC.
Oscillations are in fact a common feature in these curves, which thus go through several points
of zero net force, alternating between stable and unstable mechanical equilibria.

\begin{figure}\centering
\subfigure[]{
\includegraphics[width=7 cm]{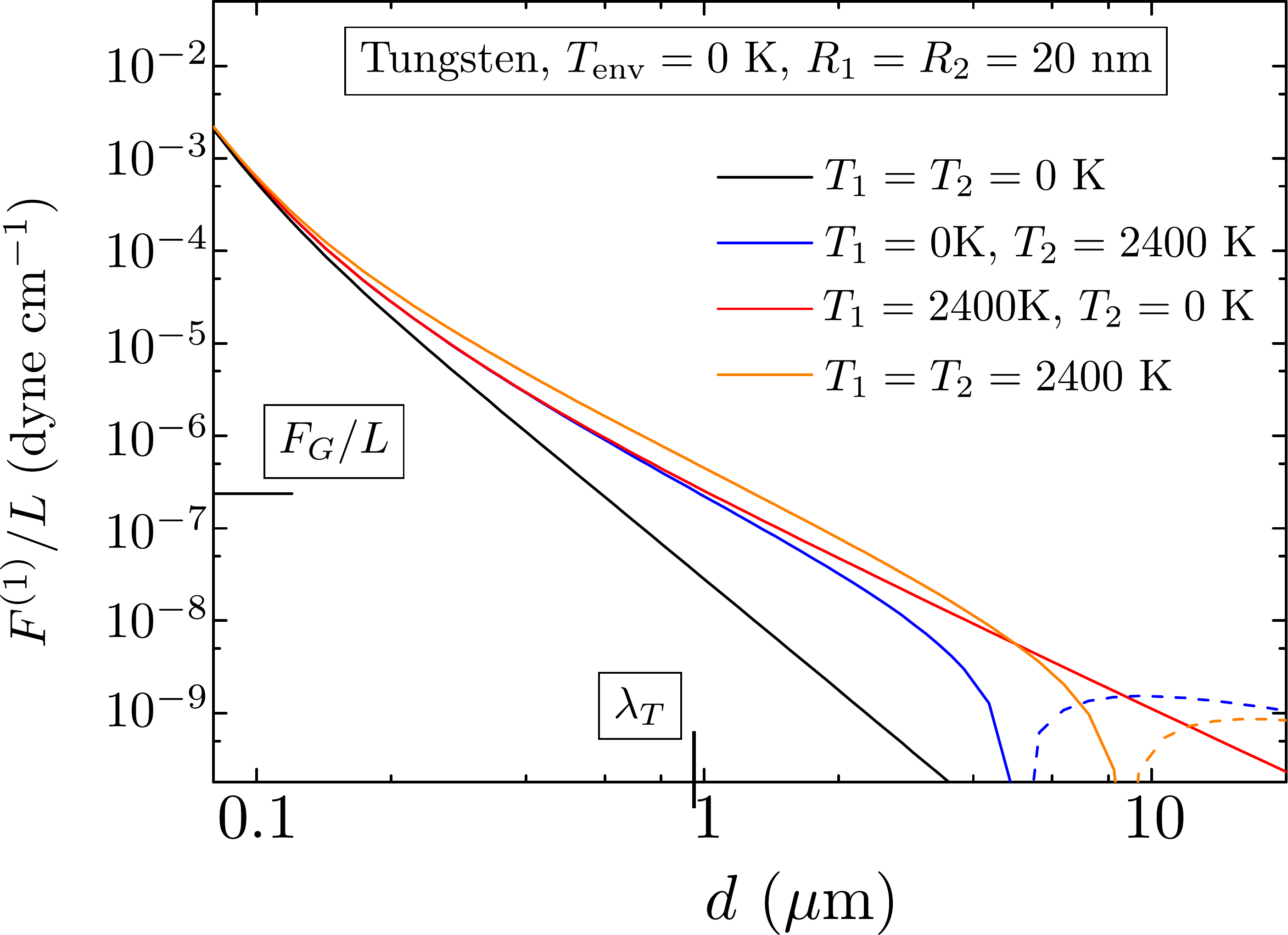}
\label{WW0}} \subfigure[]{
\includegraphics[width=7 cm]{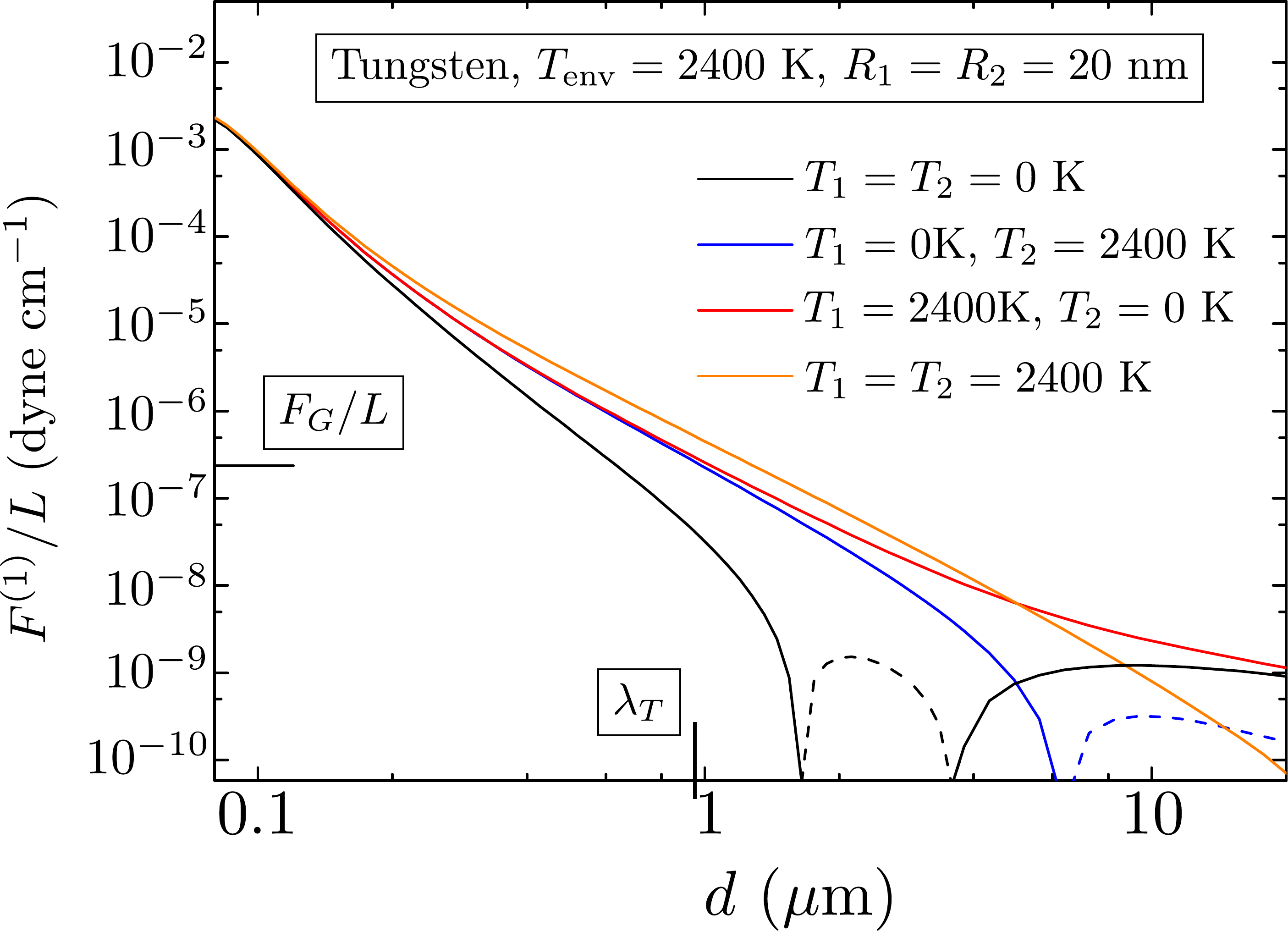}
\label{WW298}} \caption{(color online). Total force on cylinder 1 per unit length in a system of two tungsten cylinders with equal radii $R=0.02\mu$m
at separation $d$ in a a) cold (0 K) b) warm (2400K) environment. Dashed lines indicate repulsion. Horizontal line $F_G/L$ denotes the weight of the corresponding tungsten cylinder per unit length.}\label{WW}
\end{figure}

Conducting cylinders strongly radiate for radii of the order of the relevant skin-depth;
hot conducting nanowires are hence good candidates for experimental observation of thermal nonequilibrium effects.
Figure~\ref{WW} depicts the corresponding force curves for tungsten nanowires of $R_1=R_2=20$nm.
The higher temperature considered in this case is 2400K; the large temperature and small
radius are chosen for emphasis.
Indeed in this case both the equilibrium and non-equilibrium forces
are six orders of magnitude larger than for SiC (non-equilibrium effects are roughly three orders of magnitude
larger for tungsten at room temperature  compared to SiC, and additionally
enhanced by increasing the temperatures of the tubes).
The overall trends (repulsive or attractive forces) are similar to SiC, but there are fewer oscillations.
Interestingly, for conductors, the non-equilibrium correction to the Casimir forces can be two orders larger
than the weight of the cylinder, thereby being in the observable range in principle.
We also give a rough estimate of Ampere's force between the nanowires, based on the currents necessary
to keep their temperatures constant, which is much smaller than the non-equilibrium correction, rendering
conducting nanowires potential candidates for experimental studies of non-equilibrium Casimir forces.\footnote
{We note however that the validity of a continuous dielectric function when the size of the object is comparable to the mean free path of the charges  (typically a few nanometers for conductors) is an issue to be addressed in future studies. For example, plasmon resonances can occur in this case~\cite{plasmon}, which might change the non-equilibrium phenomenology in interesting ways.}

The above results (as well as input dielectric functions and other  details) for SiC and W are discussed
in Secs.~\ref{sec:SiC} and \ref{sec:W}. Additional computational details, such as the expansion of
scattering amplitudes at small cylinder radius, are relegated to Appendices.

\section{Non-equilibrium Casimir forces between cylinders}\label{Analytic}
\subsection{Formalism}\label{formalism}
We apply the formalism introduced in Ref.~\cite{mkprl} and further discussed in Ref.~\cite{mkepl},
to a system of two parallel cylinders of radii $R_j$ ($j=1,2)$ and axis-to-axis separation $d$ in vacuum.
Each object is held at a constant and homogenous temperature $\{T_j\}$, and embedded in an environment
at temperature $T_{env}$.
The cylinders are characterized by electric and magnetic responses  $\varepsilon_j$ and  $\mu_j$.
For  anisotropic materials $\varepsilon_j$ can be a tensor; we specialize to objects with azimuthal and translational
(in the direction of the cylindrical axis) symmetries such that the problem can be decomposed into
independent sectors indexed by $n$ and $k_z$ (see below).
Such a description is valid for isotropic as well as for uniaxial materials (with optical axis parallel to the cylinder axis);
such a model was recently proposed as a description of multi-walled carbon nanotubes~\cite{vgpre}.


In a non-equilibrium steady state, the net force on the system may be non-zero since there is momentum carried
away by the field radiated to the environment. Thus we need to consider the force acting on each cylinder separately.
We denote the total force acting on the cylinder 1 by $\vct{F}^{(1)}$ , whereas $\vct{F}^{(2)}$ can be found directly
from the expression for $\vct{F}^{(1)}$ by interchanging indices 1 and 2 and  changing its sign.
As shown in Refs.~\cite{mkprl,mkepl},
\begin{equation}\label{fforce}
\begin{split}
{\bf F}^{(1)}(T_{env},T_1,T_2)=&{\bf F}^{(1),eq}(T_{env})\\&+\!\sum_{j=1,2} \!\left[{\bf F}^{(1)}_{j}(T_j)- {\bf F}^{(1)}_{j}(T_{env})\right] ,
\end{split}
\end{equation}
where ${\bf F}^{(1),eq}(T_{env})$ is the Casimir force between the cylinders for the case of global equilibrium
at temperature $T_{env}$, containing the contribution from zero point fluctuations.
This force is not discussed in the present paper and treated as known (it can be computed using
scattering results for cylinders as in Refs.~\cite{Rahi08,emig11}; more general numerical schemes exist
for calculating Casimir  forces between 3D objects of arbitrary shapes and dielectric properties~\cite{homer09}).
The {\it difference} of ${\bf F}^{(1)}(T_{env},T_1,T_2)$ from ${\bf F}^{(1),eq}(T_{env})$ is due to the {\it deviations} of the cylinder temperatures $T_1$ and $T_2$ from $T_{env}$. ${\bf F}^{(1)}_{j}(T)$ is the force acting on cylinder $1$ due to
the sources in cylinders $j=1,2$ at temperature $T$. Although dealing with three sources (including the environment),
we have thus only to evaluate two terms: ${\bf F}^{(1)}_{1}(T)$ and ${\bf F}^{(1)}_{2}(T)$.
Due to  symmetry of the system, the net force is parallel to the axis-normal connecting the two axes,
and we denote its magnitude by ${ F}^{(1)}_{1}(T)$ and ${ F}^{(1)}_{2}(T)$.

Computation of forces proceeds through integrating appropriate components of the EM stress--energy tensor
around a surface enclosing one (or both) cylinders.
In the Rytov formalism~\cite{Rytov}, correlations of the electric field $\vct{E}$ at frequency $\omega$ are sourced by
sources in the two cylinders.
The contribution of each cylinder can be computed separately~\cite{vgpre}, and for a pair of points at
$\vct{r}$ and $\vct{r}'$ outside the cylinder, the symmetrized correlator has the following spectral density,
\begin{equation}\label{EXE}
\begin{split}
&\left\langle \vct{E}(t,\vct{r})\otimes\vct{E}(t',\vct{r}')\right\rangle_{\rm sym}=\int_{-\infty}^\infty \frac{d\omega}{(2\pi)^2}e^{-i\omega(t-t')}C_j(T_j,\omega),\\
&C_j(T_j,\omega)=-a(T_j,\omega)\sum_{\{P,P'\}=M,N}\sum_{n=-\infty}^{\infty}\int_{-\infty}^{\infty}\frac{dk_z}{8\pi}\\&\times A_{j,n,k_z}
\textbf{P}_{n,k_z}(\textbf{r})\otimes\textbf{P}'^*_{n,k_z}(\textbf{r}'),\\
\end{split}
\end{equation}
where $\textbf{M}_{n,k_z}$ and $\textbf{N}_{n,k_z}$ are the two polarized outgoing cylindrical waves, see App.~\ref{A}.
These waves are indexed by the multipole order $n$ and $k_z$, the component of the wavenumber $k=\omega/c$ along
the cylinder axis. The overall strength of correlations is set by
\begin{align}
a(T,\omega)= \frac{\omega^2 \hbar
  (4\pi)^2}{c^2}(\exp[\hbar \omega/k_BT]-1)^{-1}\, ,
\end{align}
which describes the occupation
number of all oscillators of frequency $\omega$; $c$ is the speed of light, and
$\hbar$ is Planck's constant. We have also introduced the abbreviation
\begin{equation}
\begin{split}
&A_{j,n,k_z}^{PP'}=\left(\textrm{Re}T_{j,n,k,z}^{PP'}+\sum_{P''=M,N}T_{j,n,k_z}^{PP''}T_{j,n,k_z}^{P'P''*}\right)\\&\times\Theta\left(\frac{\omega}{c}-|k_z|\right)+(-1)^n \textrm{Re}T_{j,n,k,z}^{PP'}\Theta\left(|k_z|-\frac{\omega}{c}\right),\\
\end{split}
\end{equation}
describing the contribution of propagating and evanescent waves. Note that only propagating waves contribute to the heat emitted by a single cylinder, whereas evanescent waves also contribute to the interactions of two cylinders.
$T_{j,n,k_z}^{PP'}$ is the $\mathbb{T}$-matrix element of cylinder $j$, which relates the amplitude of a scattered wave of polarization $P$ in response to an incoming wave of unit amplitude and polarization $P'$. The explicit form of $\mathbb{T}$-matrix elements for isotropic or uniaxial materials can  be found in Sec.~III of Ref.\cite{vgpre} (see also \cite{Bohren} for the isotropic case). In general, the $\mathbb{T}$ operator of a cylindrical object is symmetric and diagonal in $n$ and $k_z$, but couples different polarizations.

\subsection{Interaction Force}\label{Fint}
In this subsection we compute the force exerted on cylinder 1 by the field produced by cylinder 2, i.e., the interaction force ${ F}^{(1)}_{2}(T)$. To this end, we  scatter the field in Eq.~\eqref{EXE} (for $j=2$) at cylinder 1 and then
compute the force by integration of the Maxwell stress tensor on a surface enclosing 1.
Since multiple reflections on the cylinders are ignored this is a one reflection approximation~\cite{mkepl},
valid for large $d/R$, which results in a force per unit length of
\begin{equation}\label{ifgen}
\begin{split}
&\lim_{d\gg R}\frac{F^{(1)}_2}L=\frac{\hbar}{2\pi^2}\int_0^{\infty} \frac{d\omega}{e^{\frac{\hbar \omega}{k_B T_2}}-1}\sum_{P,P'}\sum_{n,m=-\infty}^{\infty}\left[(-1)^{(n+m)}\right.\\&\left.\int_{|k_z|>\omega/c}dk_z |q| \widetilde{f}^{1PP'}_{2,n,m,k_z}-\int_{|k_z|<\omega/c}dk_z q f^{1PP'}_{2,n,m,k_z}\right],\\
\end{split}
\end{equation}
where $q=\sqrt{k^2-k_z^2}$ is the wavenumber component perpendicular to the cylindrical axis.
Note that negative and positive signs of the numerical result correspond to attraction and repulsion respectively
(again, the force is parallel to the axis-to-axis separation). The functions $f$ and $\tilde{f}$ in Eq.~\eqref{ifgen} take the
rather lengthy forms
\begin{widetext}
\begin{equation}\label{fdef}
\begin{split}
&f^{1PP'}_{2,n,m,k_z}=\Re A^{PP'}_{2,n,k,z}\Im\left[H_{n-m}^{(1)}(q d)H_{n-m-1}^{(1)*}(q d)\left(T_{1,m,k_z}^{PP'}+T_{1,m+1,k_z}^{P'P*}+2\sum_{P''}T_{1,m,k_z}^{PP''}T_{1,m+1,k_z}^{P''P'*}\right)\right]\\&+2 \Im A^{PP'}_{2,n,k_z}\Re\left[H_{n-m}^{(1)}(q d)H_{n-m-1}^{(1)*}(q d)\sum_{P''}T_{1,m,k_z}^{PP''}T_{1,m+1,k_z}^{P''P'*}\right],\\
&\widetilde{f}^{1PP'}_{2,n,m,k_z}=\Re T^{PP'}_{2,n,k_z} \Re\left[H_{n-m}^{(1)}(q d)H_{n-m-1}^{(1)*}(q d)\left(T_{1,m,k_z}^{PP'}+T_{1,m+1,k_z}^{PP'*}\right)\right],\\
\end{split}
\end{equation}
\end{widetext}
where $H^{(1)}_n$ stands for the Hankel function of first kind of order $n$. The interaction force in Eq.~\eqref{ifgen} consists of two distinct terms: Due to propagating ($|k_z|<\omega/c$) and evanescent waves ($|k_z|>\omega/c$), whose properties will be discussed below. We note that for nanowires with thickness in the range of a few tens of nanometers, Eq.~\eqref{ifgen} gives accurate results down to separations of well below 1 micron.

\underline {Thin cylinders ($\{R_j\} \ll d,\{\delta_j\},\lambda_{T} $):}
The general expression in Eq.~\eqref{ifgen} is complicated as it involves an infinite series; more insightful expressions are obtained in the asymptotic limit of thin cylinders. In this paragraph we analytically study the limit $\{R_j\} \ll d,\{\delta_j\},\lambda_{T}$,  {where $\delta_j=c/\Im[\sqrt{\varepsilon_j}]\omega$ is the skin-depth of tube $j$}. The additional limits $\{R_j\}\ll\{\delta_j\},\lambda_T$ allow restriction to a finite number of partial waves whose
$\mathbb{T}$-matrix elements  are proportional to $R_j^2$ (see App.~\ref{B} and note that we take $\mu=1$).
Thus all terms with $|n|$ or $|m|$ larger than 1, and the terms with products of two $\mathbb{T}$-matrix elements, can be neglected.
The corresponding force can be found analytically in the two limits concerning the ratio of thermal wavelength and separation. For $d\ll \lambda_{T_2}$, we have
\begin{equation}\label{smalldbiggerthanRtotal}
\begin{split}
&\lim_{\{R_j\} \ll d,\{\delta_j\},\lambda_{T_2}  } \frac{F^{(1)}_2}L=\hbar\int_0^{\infty} \frac{d\omega}{e^{\frac{\hbar \omega}{k_B T_2}}-1}
R_1^2R_2^2\\&\times\left[\frac{g^{in}_6(\varepsilon_1(\omega),\varepsilon_2(\omega))} {d^6}+\frac{\omega^2 g^{in}_4(\varepsilon_1(\omega),\varepsilon_2(\omega))}{c^2d^4}\right]
,
\\
\end{split}
\end{equation}
where
 the auxiliary functions $g^{in}_6$ and $g^{in}_4$, that depend only on the dielectric functions, are given in Eqs.~\eqref{gin6} and \eqref{gin4}, respectively. We note that the force in Eq.~\eqref{smalldbiggerthanRtotal} is in most cases attractive and dominated by the evanescent part. (It can nevertheless be made repulsive for certain dielectric functions, as discussed in Ref.~\cite{bimo11}). At large separations, i.e., $d\gg \lambda_{T_2}$, the leading term of the interaction force is,
\begin{equation}\label{bigd}
\begin{split}
&\lim_{\{R_j\}\ll d,\{\delta_j\},\lambda_{T_2} } \frac{F^{(1)}_2}L=\int_0^{\infty} \frac{d\omega \omega^5 R_1^2 R_2^2}{e^{\frac{\hbar \omega}{k_B T_2}}-1}
\frac{ g^{in}_1(\varepsilon_1(\omega),\varepsilon_2(\omega))}{c^5 d}
,\\
\end{split}
\end{equation}
where $g^{in}_1$ is given in Eq.~\eqref{gin1}. This limit always yields a repulsive force which is due to momentum transfer by propagating waves. Evanescent waves do not contribute in this order as they decay too quickly with separation.

Further insight can be gained by additionally requiring the temperature to be so low that one can expand the dielectric function, i.e., $\lambda_{T_2}\gg\lambda_{0_j}$, where $\lambda_{0_j}$ is the
wavelength of the lowest resonance of the dielectric response of cylinder $j$. According to Ref.~\cite{Jackson},
\begin{equation}\label{EpsilonLowT}
\lim_{\lambda_{0_j}\ll\lambda_{T_2}}\varepsilon_{j}(\omega)=\varepsilon_{0_j}+i\frac{\lambda_{in_j}\omega}{c}+\mathcal{O}(\omega^2),
\end{equation}
with real valued static dielectric constant $\varepsilon_{0_j}$ and inelastic collision length $\lambda_{in_j}$.
With this form of $\varepsilon_{j}(\omega)$, the corresponding leading behaviors of  the force are
\begin{equation}\label{smalldbiggerthanRtotallowT}
\begin{split}
\lim_{\tiny {\begin{array}{c}
        d,\{\lambda_{0_j}\}\ll \lambda_{T_2} \\
       d, \{\delta_j\},\lambda_{T_2} \gg \{R_j\}
        \end{array}
        }} \frac{F^{(1)}_2}L=&-\frac{\hbar c \lambda_{in_2} R_1^2 R_2^2 f_6^{in}(\varepsilon_{0_1},\varepsilon_{0_2})}{\lambda_{T_2}^2 d^6 }\\&-\frac{\hbar c \lambda_{in_2} R_1^2 R_2^2 f_4^{in}(\varepsilon_{0_1},\varepsilon_{0_2})}{    \lambda_{T_2}^4 d^4},
\end{split}
\end{equation}
and
\begin{equation}\label{bigdlowT}
\begin{split}
&\lim_{\tiny {\begin{array}{c}
      d\gg\lambda_{T_2}\gg\{\lambda_{0_j}\} \\
       d, \{\delta_j\},\lambda_{T_2} \gg \{R_j\}
        \end{array}
        }}\frac{F^{(1)}_2}L=\frac{\hbar c \lambda_{in_1}\lambda_{in_2} R_1^2 R_2^2 f^{in}_1(\varepsilon_{0_1},\varepsilon_{0_2})}{\lambda_{T_2}^8 d}.\\
\end{split}
\end{equation}
Note that the dependence of temperature (via powers of $\lambda_T$) is quite different in these limits.
The corresponding auxiliary functions $f^{in}_6$, $f^{in}_4 $ and $f^{in}_1$ are given by Eqs.~\eqref{fin6}, \eqref{fin4} and \eqref{fin1}, respectively.

\subsection{Self-force}\label{Fself}
We next compute the self force $F^{(1)}_1$, acting on cylinder 1 due to the field emitted by cylinder 1 and reflected from cylinder 2, again restricting  to  $d\gg\{R_j\}$, where the one reflection approximation is valid.
The origin of this force is cylinder 2 acting as a reflector of EM waves emitted by cylinder 1.
The interference of the emitted and reflected waves is expected to give rise to the observed oscillatory behavior of the self-force as a function of separation.
Computationally it is easier to first evaluate the force $F_1^{(1+2)}=F^{(2)}_1+F^{(1)}_1$ acting on both cylinders due to  sources on cylinder 1, and to then subtract $F^{(2)}_1$.
Again, it is a property of non-equilibrium that the net force on the system is not  zero.
Note that $F^{(2)}_1$ is obtained from Eq.~\eqref{ifgen} by interchanging indices 1 and 2 and changing its overall sign.

After tedious algebra we arrive at the following expression for the force $F_1^{(1+2)}$,
\begin{equation}\label{selfforce}
\begin{split}
&\lim_{d\gg R}\frac{F_1^{(1+2)}}L=\frac{\hbar}{2\pi^2}\int_0^{\infty} \frac{d\omega}{e^{\frac{\hbar \omega}{k_B T_1}}-1}\\&\sum_{\{P,P'\}=N,M}\sum_{n,m=-\infty}^{\infty}\int_{|k_z|<\omega/c}dk_z
 q s^{PP'}_{1,n,m,k_z},\\
\end{split}
\end{equation}
where
\begin{equation}\label{forcecomps}
\begin{split}
&s^{PP'}_{1,n,m,k_z}=2 \Re A^{PP'}_{1,n,k_z} \Im\left[H_{n-m}^{(1)}(q d)J_{n-m-1}(q d)T_{2,m,k_z}^{PP'}
\right.\\&\left.+J_{n-m}(q d)H_{n-m-1}^{(1)*}(q d) T_{2,m+1,k_z}^{PP'*}\right],\\
\end{split}
\end{equation}
and $J_n$ denotes the Bessel function of order $n$.
As expected, the force on the system is solely due to propagating waves, since evanescent waves do not carry momentum to the environment.

\underline {Thin cylinders ($\{R_j\} \ll d,\{\delta_j\},\lambda_{T} $):} We gain further insight by examining $F_1^{(1+2)}$ in the asymptotic limit of thin dielectric cylinders.
For $d\ll \lambda_{T_1}$,
\begin{equation}\label{smalldtotforce}
\begin{split}
&\lim_{d,\{\delta_j\},\lambda_{T_1} \gg \{R_j\} } \frac{F_{1}^{(1+2)}}L=\mathcal{O}(d^{-1})
\end{split}
\end{equation}
i.e. diverging weakly at small $d$. Compared to this, the force $F^{(2)}_1$ in Eq.~\eqref{smalldbiggerthanRtotal} is proportional to $d^{-6}$, and we can neglect $F_1^{(1+2)}$ in this regime, so that for $d\ll \lambda_{T_1}$
$$\lim_{ d,  \{\delta_j\},\lambda_{T_1} \gg \{R_j\} }F_1^{(1)}=-\lim_{d,
  \{\delta_j\},\lambda_{T_1} \gg \{R_j\} } F^{(2)}_1.$$
We thus observe that at close separations these contributions to the forces on the cylinders are equal and opposite,
again because  momentum transfer to the environment can be ignored.

In the opposite limit of large distances, $d\gg \lambda_{T_1}$, one can verify that
\begin{equation}
\begin{split}\label{bigdtotalforce}
&\lim_{ d,  \{\delta_j\},\lambda_{T_1} \gg \{R_j\} }\frac{F^{(1)}_1}L =\mathcal{O}({d^{-3/2}}),
\end{split}
\end{equation}
where the prefactor is a lengthy algebraic function additionally depending on $d$ in an oscillatory manner via $\exp{(2i\omega d/c)}$. Thus, at large separations the self-force is asymptotically less relevant compared to $F^{(1)}_2$ (following ${d^{-1}}$ in Eq.~\eqref{bigd}).

\subsection{Cylinders from spheres in the dilute limit}\label{sec:sphere}\label{dilute}
A rarified material can be regarded as a collection of independent molecules; forces between two
such materials may then be obtained by pairwise summation of forces between their molecules.
Indeed, equilibrium Casimir forces in the optically dilute limit ($\varepsilon_j(\omega)\to 1$)
can be calculated by integrating the interactions between volume elements.
In this subsection we show that even the non-equilibrium force between two cylinders can be obtained
from the non-equilibrium force between two spheres~\cite{mkepl}, by appropriate summation
in the optically dilute limit.

\begin{figure}\centering
\includegraphics[width=7.0 cm]{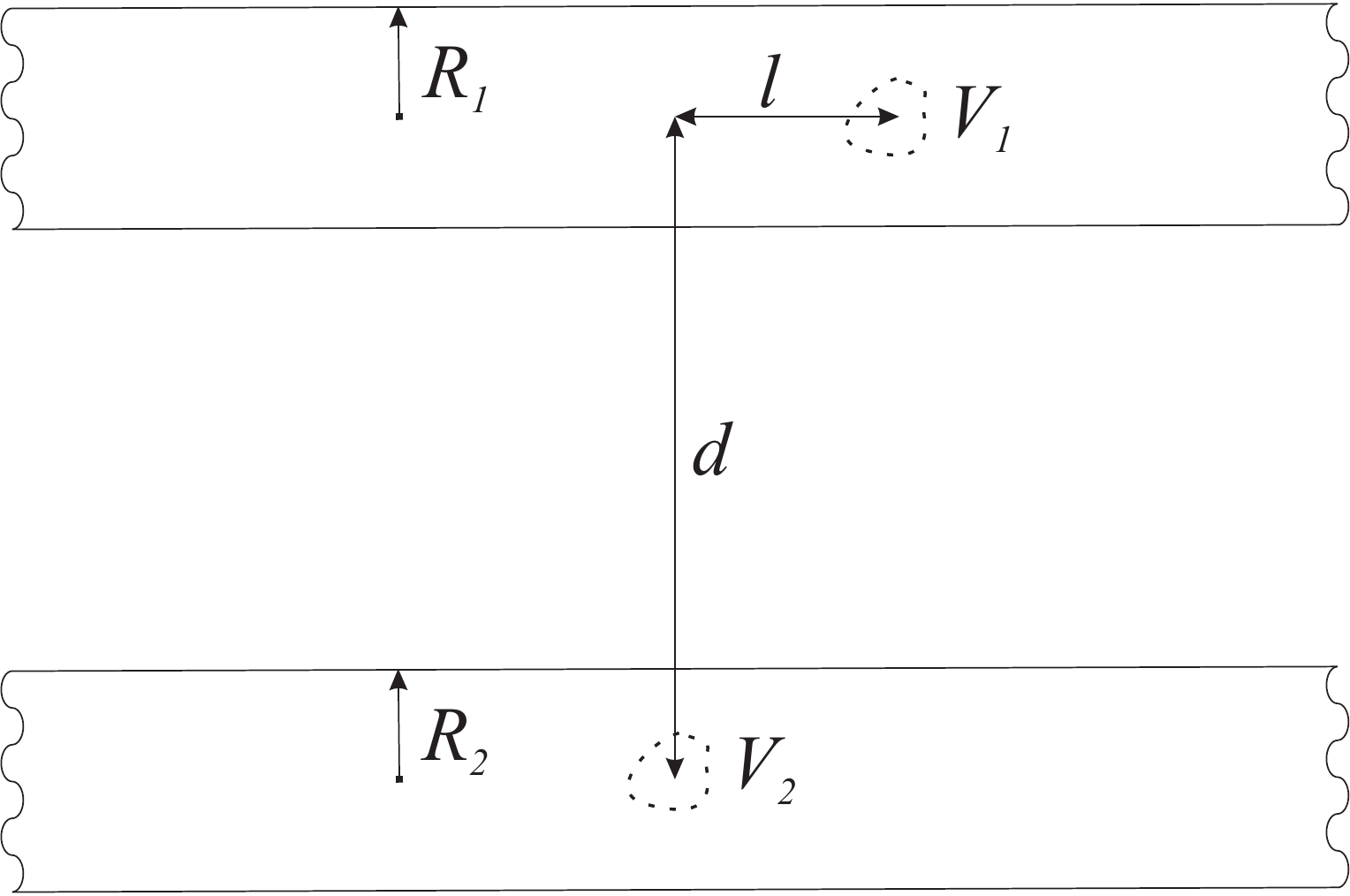}
\caption{Casimir forces between objects can also be obtained by pairwise summation of the forces between all the volume elements.}\label{dilute1}
\end{figure}

We start with $F^{(1)}_2$, where we get from Eqs.~\eqref{smalldbiggerthanRtotal} and \eqref{bigd} in the limit of $\varepsilon_{j}\to 1$, (note that this  means that $\delta_j$ is much larger than the radius of the objects)
\begin{equation}\label{cyldil}
\begin{split}
&\lim_{
        \lambda_{T_2}\gg d \gg \{R_j\}
        } \frac{F^{(1)}_2}L=-\hbar\int_0^{\infty} \frac{d\omega}{e^{\frac{\hbar \omega}{k_B T_2}}-1} R_1^2R_2^2 \\&\times \Re[\varepsilon_1-1] \Im[\varepsilon_2]\left(\frac{45}{64d^6}+\frac{3\omega^2}{16c^2d^4}\right),
\end{split}
\end{equation}
and
\begin{equation}\label{cyldilbigd}
\begin{split}
&\lim_{  d\gg \lambda_{T_2} \gg \{R_j\}}\frac{F^{(1)}_2}L=\frac{\hbar}{2\pi}\int_0^{\infty} \frac{d\omega}{e^{\frac{\hbar \omega}{k_B T_2}}-1}R_1^2R_2^2\\&\times \frac{\omega^5 }{c^5 d} \Im[\varepsilon_1] \Im[\varepsilon_2].\\
\end{split}
\end{equation}
On the other hand, Eq. (6) from Ref.~\cite{mkepl} for the interaction force between two spheres can be expanded in the dilute limit to yield the force between two volume elements $V_1$ and $V_2$,
\footnote{We have an additional minus sign compared to Eq.~(6) from Ref.~\cite{mkepl}, and we exchanged indices $1\leftrightarrow2$.}
\begin{equation}\label{eq6}
\begin{split}
&\lim_{
        d,\lambda_{T_2} \gg \{R_j\}
        } \mathcal{F}^{(1)}_2(d)=\frac{V_1V_2\hbar}{4c^7\pi^3}\int_0^{\infty} \frac{\omega^7 d\omega}{e^{\frac{\hbar \omega}{k_B T_2}}-1} \Im[\varepsilon_2]\\
        &
        \left[\frac{c^2}{\omega^2d^2}\Im[\varepsilon_1]
        -\Re[\varepsilon_1-1]
        \left(\frac{c^3}{\omega^3d^3}+\frac{2c^5}{\omega^5d^5}+\frac{9c^7}{\omega^7d^7}
        \right)
        \right],\\
\end{split}
\end{equation}
The force between two cylinders follows then as an integral over the volume of the two cylinders, see Fig.~\ref{dilute1},
\begin{equation}
\frac{F^{(1)}_2}{L}=\pi^2R_1^2R_2^2\int_{-\infty}^\infty dl\frac{\mathcal{F}^{(1)}_2(\sqrt{d^2+l^2})}{V_1V_2}\frac{d}{\sqrt{d^2+l^2}}.
\end{equation}
Using this and Eq.~\eqref{eq6}, the terms in Eqs.~\eqref{cyldil} and \eqref{cyldilbigd} are recovered. Since the consistency of the two results is physically expected, it provides a useful check for both calculations.
We note, however, that we could not recover the term proportional to $d^{-2}$ in $F^{(1)}_2$ (corresponding to $d^{-3}$ in Eq.~\eqref{eq6}) due to mathematical difficulties in the integral over special cylindrical functions.

We note also the corresponding equivalence for the self force, though somewhat more involved mathematically: In Eq.~(7) from Ref.~\cite{mkepl}, the self force decays as  $d^{-2}$ with an oscillating prefactor $\exp{(2 i \omega d/c)}$. Pairwise summation of this result yields exactly a $d^{-3/2}$ separation dependence modulated by $\exp{(2 i \omega d/c)}$, as discussed in relation to Eq.~\eqref{bigdtotalforce}. Moreover, if we represent a plate of finite thickness by an infinite set of cylinders and place a small sphere in front of this plate, the self-force acting on the sphere scales as $d^{-1}$ with the prefactor of $\exp{(2 i \omega d/c)}$. This matches perfectly with the result for the sphere/plate geometry, in Eq.~(19) from Ref.~\cite{mkepl}, in the limit of large separations.
Thus, in the dilute limit one regains the expected connections between different geometries.

\section{Numerical Examples}\label{numerics}
In this section we numerically compute the total Casimir force between two  cylinders,
in one case composed of dielectrics (SiC) and of conductors (tungsten) in the other.
The equilibrium contributions to the force (see Eq.~\eqref{fforce}) were computed using
the methods of Rec.~\cite{homer09} as implemented in the {\sc scuff-em} code suite~\cite{homercode}.

\subsection{Silicon carbide}\label{sec:SiC}

Figure~\ref{SiC0} depicts the forces on SiC cylinders of radii $R_1=R_2=0.1\mu$m in both cold (0 K) and warm (300K) environments. The optical properties of SiC are modeled by the dielectric function~\cite{ChenNew},
\begin{equation*}
\varepsilon_{SiC}(\omega)=\varepsilon_\infty\frac{\omega^2-\omega_{LO}^2+i\omega\gamma}{\omega^2-\omega_{TO}^2+i\omega\gamma},
\end{equation*}
where $\varepsilon_\infty=6.7$, $\omega_{LO}=0.12$eV, $\omega_{TO}=0.098$eV, $\gamma=5.88\times10^{-4}$eV.
We evaluated Eqs.~\eqref{ifgen} and \eqref{selfforce} numerically, restricting $n$ and $m$ to orders $\{-1,0,1\}$, and omitting quadratic terms in $\mathbb{T}$, which is justified for $\{R_j\}\ll\{\lambda_{T_j}\},\{\delta_j\}$, where the forces per unit length will be proportional to $R_1^2R_2^2$. We note that $R_j=0.1 \mu$m is roughly an upper bound for the validity of this asymptotic behavior, as we have checked by including higher order terms. Due to the small thickness of the cylinders, the resulting phenomenology is in close analogy to the case of two spheres~\cite{mkepl} (compare also Sec.~\ref{dilute}):
In the case when $T_{env}=$0K, the force starts to deviate strongly from its equilibrium value around $d\approx \lambda_T/2$, where $\lambda_T=\frac{\hbar c}{k_BT}\approx 7.6\mu$m. Cylinder 1 is repelled at large $d$ if $T_2=300$K due to  radiation pressure with a force that decays with distance as $d^{-1}$. On the other hand, if additionally $T_1=300$K, the oscillating force $F^{(1)}_1$ is appreciable and in fact dominates the total force for large $d$ if $T_1=0 K$; the net force now has many zero crossings, where every second one is a {\it stable point of zero force}. The wavelength of the oscillations is roughly $6$~$\mu$m due to the optical resonance of SiC at wavelength $\lambda_{0}\approx12\mu$m (the length of the optical path from cylinder 1 to cylinder 2 and back is $2d$, and for a sharp resonance of the dielectric function of cylinder 1 at $\lambda_{0}$, we have constructive interference at $2d=\lambda_{0}, 2\lambda_{0},\dots$, for an oscillation wavelength of $\lambda_0/2$).
Note that this figure also provides complete information about the force on cylinder 2: e.g., in case $\{T_1=0, T_2=300{\rm K}\}$, the blue curve shows the force acting on cylinder 1, while the red one represents the force on cylinder 2. At the crossing of the solid red and dashed blue curves the two cylinders feel equal forces in {\it the same direction}, an effect which might have less practical importance for tubes compared to spheres~\cite{mkepl}.

\subsection{Tungsten}\label{sec:W}

Let us turn to conductors at high temperatures. Figure~\ref{WW} illustrates the total Casimir force for two very thin tungsten cylinders (nanowires) of equal radii $R_j=0.02\mu$m, and large temperature differences of 2400K. We have chosen this value for $R_j$ as it corresponds to the maximum in emissivity of an isolated tungsten cylinder~\cite{vgpre} (which is connected to the skin depth being of the same order). As the non-equilibrium force strongly depends on the heat radiation of the objects, we expect it to be also comparatively large for this value. Also, tungsten has a relatively high melting temperature, and the value of 2400K promises large effects. Note that Eqs.~\eqref{smalldbiggerthanRtotal} and \eqref{bigd} are not valid here, and the force is not proportional to $R_1^2R_2^2$, despite the small thickness of the wires.

The following dielectric function for tungsten was used~\cite{Wdiel},
\begin{equation}\label{Tdiel}
\varepsilon_W(\omega)=1
-\frac{\lambda^2}{2\pi c\epsilon_0}\sum_{q=1}^2\frac{
\sigma_q}{\lambda_{rq}-i\lambda},
\end{equation}
where $\lambda$ is the wavelength in vacuum, $c$ is the speed of light and $\epsilon_0$ is the permittivity of vacuum in SI units.
The remaining parameters are : $\sigma_1=1.19\times 10^6$ ohm$^{-1}$m$^{-1}$, $\sigma_2=0.25\times 10^6$ ohm$^{-1}$m$^{-1}$, $\lambda_{r1}=3.66 \mu$m, $\lambda_{r2}=0.36 \mu$m. These values were obtained by fitting the dielectric function of tungsten at $T=2400$K~\cite{Wdiel}~\footnote{For the accurate evaluation of the total Casimir force with Eq.~\eqref{fforce}, one must use the cylinder dielectric function at its corresponding temperature. Thus,  Eq.~\eqref{Tdiel} is strictly only appropriate for $T_1=T_2=2400$K in Fig.~\ref{WW}. However, the dielectric function does not depend significantly on $T$ and the curves for $T_j=0$K in Fig.~\ref{WW} are good estimates.}.

The relative deviation of the total Casimir force from its equilibrium form is not as pronounced as in Fig.\ref{SiC},
since equilibrium forces between conductors decay much slower with distance compared to dielectrics.
{In particular, a Drude dielectric function gives rise to an equilibrium force scaling as $d^{-4}/\log{(d/R)}$~\cite{emig11},
in good agreement with our numerical results.
(For dielectrics, our numerical result suggests a $d^{-7}$ law of the equilibrium Casimir force in the small separation retarded regime)}.
Nevertheless, as in the dielectric case, the force in Fig.\ref{WW}  starts to deviate from the equilibrium curve
at approximately $\lambda_T/2$ with a corresponding $\lambda_T\approx 0.95\mu$m.
In a cold environment, $T_{env}=0K$, the force is attractive at small separations, whereas at large distances it becomes repulsive if $T_2=2400$K. When we increase the environment temperature to $T_{env}=2400$K, the force shows another feature for $T_1=T_2=0K$, where it is attractive for both small and large distances, but for an intermediate region it becomes repulsive yielding a stable point of zero force in the vicinity of $d\approx4\mu$m. This intermediate repulsion is due to the contribution from evanescent waves (which is repulsive because $F^{(1)}_j$ enters Eq.~\eqref{fforce} with a minus sign), and then for higher $d$ propagating waves make the force attractive again. In contrast to the dielectric displayed in Fig.~\ref{SiC300}, no  oscillations are visible in Fig.~\ref{WW}. We attribute this to the absence of sharp peaks in the dielectric function of tungsten.


 {The non-equilibrium  force in Fig.~\ref{WW} is much larger compared to Fig.~\ref{SiC}, again, as a consequence of both replacing the dielectric materials with conductors, as well as choosing the higher temperature $T=2400$K instead of $T=300$K. At the point where the total Casimir force deviates from the equilibrium  {force} the magnitude of the force is roughly $10^6$ times larger for tungsten compared to SiC-- approximately $5\times10^6$ larger per unit mass.
For tungsten nanowires at $T=300$K (we do not provide the plot for this case) the non-equilibrium effects are still three orders of magnitude larger when compared to SiC. Thus, the strong enhancement is a  result of both material properties and high temperatures. Furthermore, considering wires of length $1\mu$m at high temperature $T=2400$K, the force at the point when non-equilibrium forces start to dominate is approximately $10$fN (see Fig.~\ref{WW}), significantly larger than the weight $F_G\approx0.24$fN (also depicted in the figure) of the nanowires.}

There is another force we can use for comparison: If an electric current is used to heat the wire~\cite{singer,bimonterad},
one requires a current of $17\mu$A to maintain the temperature $T=2400$K for tungsten nanowires  (obtained by equating heat losses described by Joule's law to the heat radiation for a signle cylinder predicted in Ref.~\cite{vgpre}). Such currents correspond to an Ampere's force of about $0.15$fN for  {$d=0.4\mu$m}, which is two orders of magnitude smaller than the nonequilibrium correction to the Casimir force~\footnote{This estimate neglects possible couplings between AC or DC currents and thermal current fluctuations~\cite{shapiro}}. Thus, when switching on an AC or DC current in the wires,
the change in force between them due to heating-induced non-equilibrium Casimir force is much larger than that
due to Ampere's force.
We thus conclude that hot conducting nanowires are promising candidates for measuring or using non-equilibrium Casimir effects.

\begin{acknowledgments}
We thank R.~L.~Jaffe, T.~Emig, G.~Bimonte, M.~F.~Maghrebi, and N.~Graham for helpful discussions.
This research was supported by the NSF Grant No.  DMR-08-03315, DOE grant No.  DE-FG02-02ER45977,
and the DFG grant No. KR 3844/1-1.
\end{acknowledgments}
\numberwithin{equation}{section}
\appendix
\section{Cylindrical harmonics}\label{A}
Following Ref.~\cite{Tsang}, the EM cylindrical harmonics can be written as,
$$\textbf{M}_{n,k_z}(\textbf{r})=\left[\frac{in}{qr}H^{(1)}_n(qr)\mathbf{e}_r-H_n'^{(1)}(qr)\mathbf{e}_\phi\right]e^{ik_zz+in\phi},$$
\begin{equation}\label{cylindricalharmonics}
\begin{split}
\textbf{N}_{n,k_z}(\textbf{r})&=\frac c{\omega}\left[ik_zH_n'^{(1)}(qr)\mathbf{e}_r-\frac{nk_z}{qr}H^{(1)}_n(qr)\mathbf{e}_\phi\right.\\&\left.+qH^{(1)}_n(qr)\mathbf{e}_z\right]e^{ik_zz+in\phi},\\
\end{split}
\end{equation}
where $H_n$ is the Hankel function of the first kind of order $n$.
$\textbf{M}_{n,k_z}$ and $\textbf{N}_{n,k_z}$ correspond
to outgoing magnetic multipole (TE) and electric multipole (TM) waves
respectively. Also, $k_z$ and $q$ are the wavevectors parallel and
perpendicular to the cylindrical $z$-axis respectively, satisfying
the relation $q=\sqrt{k^2-k_z^2}$, $ k=\omega/c$.
$H_n'^{(1)}$ corresponds to the first derivative with respect to the
argument. Furthermore, we denote the corresponding regular waves by
$\textbf{RM}_{n,k_z}$ and $\textbf{RN}_{n,k_z}$, which differ from
regular ones by replacing $H_n^{(1)}$ with the Bessel function $J_n$.

The above solutions correspond to transverse waves, i.e.
$\mathbf{\nabla}\cdot\mathbf{M}_{n,k_z}=\mathbf{\nabla}\cdot\mathbf{N}_{n,k_z}=0$.
Moreover, they obey useful relations
$\mathbf{M}_{n,k_z}=\frac{c}{\omega}\mathbf{\nabla}\times\mathbf{N}_{n,k_z}$,
$\mathbf{N}_{n,k_z}=\frac{c}{\omega}\mathbf{\nabla}\times\mathbf{M}_{n,k_z}$.
These relations are also valid for outgoing waves.

\section{Small $R$ expansion of the $\mathbb{T}$ operator of the cylinder}\label{B}
In order to derive Eqs.~\eqref{smalldbiggerthanRtotal},~\eqref{bigd},~\eqref{smalldtotforce} and \eqref{bigdtotalforce} we need the expansion of the $\mathbb{T}$ operator in terms of $\omega R/c$. For a cylinder made of isotropic material with magnetic permeability $\mu(\omega)$ and dielectric permittivity $\varepsilon(\omega)$, we find for the limit $R\ll\{\delta, c/\omega\}$~\cite{vgpre},
\begin{equation}\label{TdielectricTEE0}
T_{0,k_z}^{NN}=-\frac{i\pi}{4}(\varepsilon-1)(\widetilde{k}_z^2-1)(\omega
R/c)^2, \\
\end{equation}
\begin{equation}\label{TdielectricTMM0}
T_{0,k_z}^{MM}=-\frac{i\pi}{4}(\mu-1)(\widetilde{k}_z^2-1)(\omega
R/c)^2, \\
\end{equation}
\begin{equation}\label{TdielectricTEE1}
\begin{split}
T_{1,k_z}^{NN}=T_{-1,k_z}^{NN}&=\frac{i\pi}{4}\frac{\widetilde{k}_z^2(\mu+1)(\varepsilon-1)+(\mu-1)(\varepsilon+1)}{(\varepsilon+1)(\mu+1)}\\
 &\times(\omega
R/c)^2, \\
\end{split}
\end{equation}
\begin{equation}\label{TdielectricTMM1}
\begin{split}
T_{1,k_z}^{MM}=T_{-1,k_z}^{MM}&=\frac{i\pi}{4}\frac{\widetilde{k}_z^2(\mu-1)(\varepsilon+1)+(\mu+1)(\varepsilon-1)}{(\varepsilon+1)(\mu+1)}\\&\times (\omega
R/c)^2, \\
\end{split}
\end{equation}
\begin{equation}\label{TdielectricTME1}
\begin{split}
T_{1,k_z}^{MN}=T_{1,k_z}^{NM}=-T_{-1,k_z}^{MN}=-T_{-1,k_z}^{NM}&=\frac{i\pi}{2}\frac{(\varepsilon\mu-1)
\widetilde{k}_z}{(\varepsilon+1)(\mu+1)}\\&\times (\omega
R/c)^2, \\
\end{split}
\end{equation}
where $\widetilde{k}_z=k_z/k$.

\section{Auxiliary functions}\label{C}
In defining functions below, the superscript emphasizes that we deal with the interaction force, whereas the numerical subscript indicates the force's power of decay in the axis-to-axis separation between cylinders.
\begin{equation}\label{gin6}
\begin{split}
&g^{in}_6(\varepsilon_1(\omega),\varepsilon_2(\omega))=\frac{45}{2048}\Im\left[\frac1{\varepsilon_2+1}\right]\frac 1 {{|\varepsilon_1+1|^2}}\\&\times\left[(|\varepsilon_1|^2-1)(4 (33+5 \Re[\varepsilon_1])+(7+3 \Re[\varepsilon_1]) |\varepsilon_2+1|^2)\right.\\&\left.+(\Re[\varepsilon_1]^2-1)(40+6 |\varepsilon_2+1|^2)\right];
\end{split}
\end{equation}
\begin{equation}\label{gin4}
\begin{split}
&g^{in}_4(\varepsilon_1,\varepsilon_2)=\frac{3}{256}\Im\left[\frac1{\varepsilon_2+1}\right]\frac{1}{{|\varepsilon_1+1|^2}}\\&\times\left[\Im[\varepsilon_1]^2 (|\varepsilon_2+1|^2(7-\Re[\varepsilon_1])+12\Re[\varepsilon_1]+76 )\right.\\&\left.+(\Re[\varepsilon_1]^2-1)(|\varepsilon_2+1|^2(5-\Re[\varepsilon_1])+12\Re[\varepsilon_1]+100)\right];
\end{split}
\end{equation}
\begin{equation}\label{gin1}
\begin{split}
&g^{in}_1(\varepsilon_1(\omega),\varepsilon_2(\omega))= \frac 2{15\pi} \Im\left[\frac1{\varepsilon_1+1}\right]\Im\left[\frac1{\varepsilon_2+1}\right]\\&\times\left( |\varepsilon_1+1|^2 |\varepsilon_2+1|^2+|\varepsilon_1+1|^2+|\varepsilon_2+1|^2+36\right);
\end{split}
\end{equation}
\begin{equation}\label{fin6}
\begin{split}
&f_6^{in}(\varepsilon_{0_1},\varepsilon_{0_2})=\frac{15\pi^2}{4096}\\& \times\frac{(\varepsilon_{0_1}-1)\left[172+ (13+3\varepsilon_{0_1})(\varepsilon_{0_2}+1)^2+ 20 \varepsilon_{0_1}\right]}{(\varepsilon_{0_1}+1)(\varepsilon_{0_2}+1)^2};
\end{split}
\end{equation}
\begin{equation}\label{fin4}
\begin{split}
f_4^{in}(\varepsilon_{0_1},\varepsilon_{0_2})=\frac{\pi^4(\varepsilon_{0_1}-1)}{1280 (\varepsilon_{0_1}+1)}\left[\frac{12 \varepsilon_{0_1}+100}{(\varepsilon_{0_2}+1)^2}-\varepsilon_{0_1}+5\right];
\end{split}
\end{equation}
\begin{equation}\label{fin1}
\begin{split}
&f_1^{in}(\varepsilon_{0_1},\varepsilon_{0_2})=\frac{16\pi^7}{225 } \\&\times \frac{(\varepsilon_{0_1}+1)^2 + (\varepsilon_{0_2}+1)^2 + (\varepsilon_{0_1}+1)^2 (\varepsilon_{0_2}+1)^2+36}{(\varepsilon_{0_2}+1)^2(\varepsilon_{0_1}+1)^2}.
\end{split}
\end{equation}
\vspace{ 5.0cm}
%

\end{document}